\documentclass[twocolumn,showpacs,preprintnumbers,amsmath,amssymb]{revtex4} 

\usepackage{graphicx,color}
\usepackage{dcolumn}
\usepackage{bm}

\begin{document}




\title{Topological jamming of spontaneously knotted polyelectrolyte chains driven through a nanopore}

\author{A. Rosa$^1$}
\email{anrosa@sissa.it}
\author{M. Di Ventra$^2$}
\author{C. Micheletti$^{1,3}$}
\email{michelet@sissa.it}
\affiliation{$^1$ SISSA - Scuola Internazionale Superiore di Studi Avanzati, Via Bonomea 265, 34136 Trieste (Italy)\\
$^2$ Department of Physics, University of California, San Diego, La Jolla, CA 92093-0319\\
$^3$ CNR-IOM Democritos, Via Bonomea 265, 34136 Trieste (Italy)}

\date{\today}

\begin{abstract}
The advent of solid state nanodevices allows for interrogating the
physico-chemical properties of a polyelectrolyte chain by
electrophoretically driving it through a nanopore.  Salient dynamical
aspects of the translocation process have been recently characterized by
theoretical and computational studies of model polymer chains free from
self-entanglement.  However, sufficiently long equilibrated chains are
necessarily knotted. The impact of such topological ``defects'' on the
translocation process is largely unexplored, and is addressed in this
study. By using Brownian dynamics simulations on a coarse-grained
polyelectrolyte model we show that knots, despite being trapped at the
pore entrance, do not {\it per se} cause the translocation process to
jam. \textcolor{black}{Rather, knots introduce an effective friction that increases with the
applied force, and practically halts the translocation above a threshold force.} The
predicted dynamical crossover, which is experimentally verifiable, is of
relevance in applicative contexts, such as DNA nanopore sequencing.
\end{abstract}

\maketitle

Nanopores, namely holes of nanoscale dimensions carved out of biological
or solid-state membranes, are increasingly becoming an important tool to
probe chemical and physical properties of polymers
\cite{zwolak_diventra,branton08,austin_frey,dekker_joanny,Luan:2010:Soft-Matter:20563230,Electrophoretic_DNA_Translocation}.
For instance, polyelectrolytes, such as DNA, can be electrophoretically
translocated through a pore and their chemical composition can be
inferred either by the ionic current blockaded by the DNA strand
\cite{Kasianowicz1996-1}, or from the electrical current measured
perpendicular to the DNA backbone \cite{Lagerqvist2006-1}.  These
approaches are being actively investigated because they hold great
promise for fast and low-cost DNA sequencing.

However, one of the issues that has received much less attention is
related to possible limitations arising from the maximum length of the
polymer that can be electrophoretically translocated through a nanopore
without obstruction from the inevitable chain self-entanglement (knots).
In fact, it is well known that, the incidence of knots increases
exponentially with the chain contour length
\cite{Sumners&Whittington:1988:J-Phys-A}, and, in turn,
\textcolor{black}{can affect kinetic, mechanical and equilibrium
  properties of sufficiently long (bio-)polymers
  \cite{Ercolini:2007:PRL,Marenduzzo_et_al_2010_JoP_Cond_Matt,rybenkov_cozzarelli,saitta_klein,chan_zechiedrich,Deguchi&Tsurusaki:1993:J-Phys-Soc-Jap,Bao:2003:Phys-Rev-Lett:14754067,Sulkowska:2008:Phys-Rev-Lett:18352439,Huang:2008:J-Chem-Phys:19044999,Meluzzi:2010:Annu-Rev-Biophys:20192771}.}
A knot in the polymer chain is then an unwanted potential obstruction to
its translocation through a pore; much like the knot we customarily make
at the end of a sewing thread to prevent it from ``translocating''
through a threaded piece of cloth.

Motivated by these observations, in this Letter we set out to
investigate theoretically and numerically the dynamics of pore
translocation of chains whose contour length exceeds by orders of
magnitude their persistence length.  On the one hand, this allows us to
push to unprecedented contour lengths the assessment of the validity of
previously-suggested dynamical scaling relationships for the
translocation process of unknotted chains (see, e.g.,
Ref.~\cite{PhysRevE.69.021806,PhysRevE.78.021129,grosberg_transloc,sakauePRE,sakaue_raphael,forrey_muthukumar,PhysRevE.85.051803}).
\textcolor{black}{On the other hand, we can clarify the impact of spontaneous knotting on the driven translocation of biopolymers. This avenue appears to be largely unexplored except for the recent protein-related investigation of Huang and Makarov \cite{Huang:2008:J-Chem-Phys:19044999}.}

In particular, we show that knots do not {\it per se} cause the
translocation process to halt. More precisely, they are found to act as
plug-like obstructions of the pore only if a threshold driving force is
exceeded. Based on this result it is expected that accounting for the
topology-dependent dynamical crossover ought to be important for
applications that employ nanopores, such as the detection and sequencing
of DNA filaments \cite{zwolak_diventra,branton08}.

\begin{figure}
\includegraphics[width=3.0in]{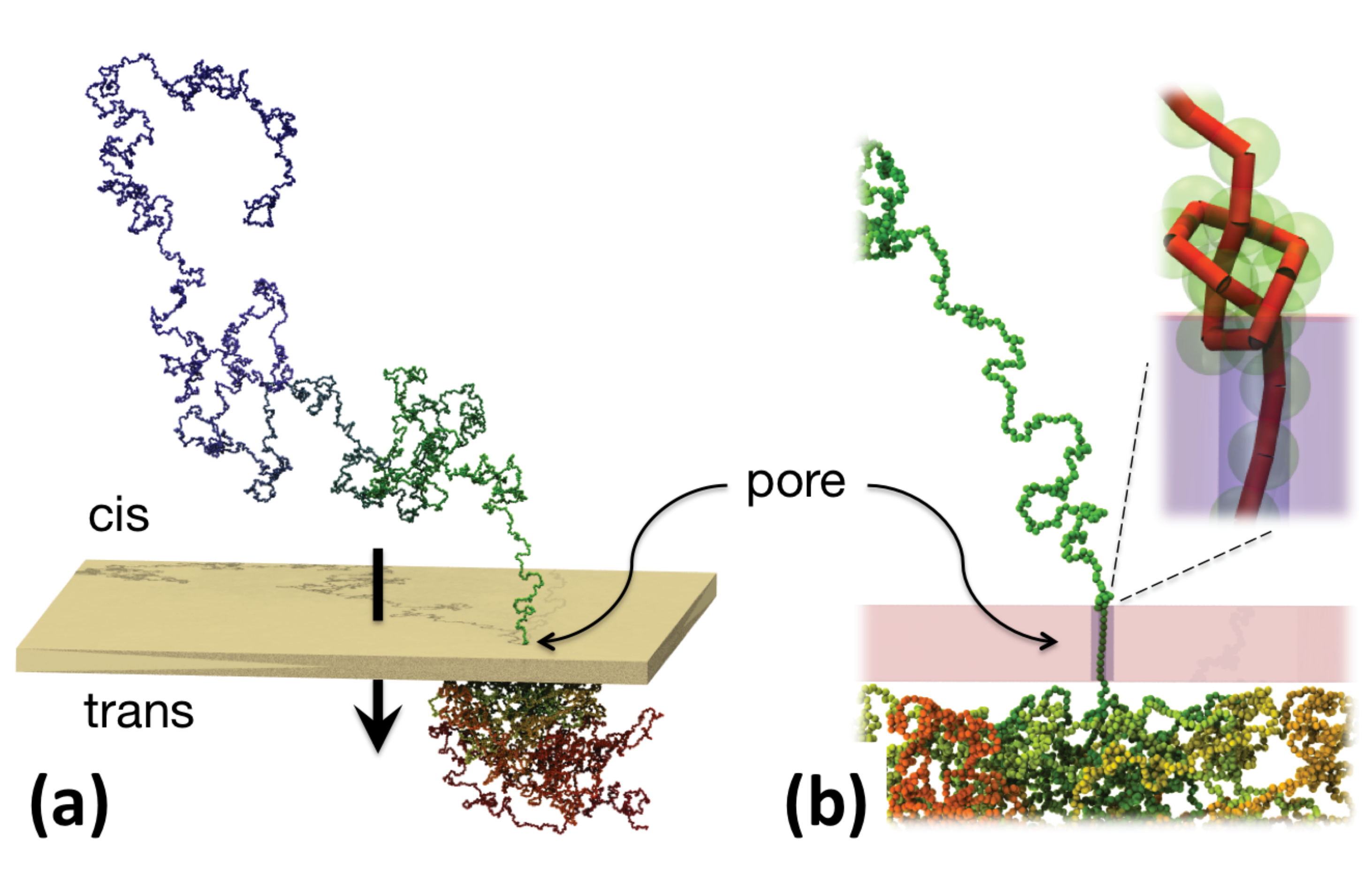}
\caption{
\label{fig:setup}
[Color online]
(a) Snapshot of an instantaneous configuration of the  $15 \mu$m-long
model polyelectrolyte driven electrophoretically through the nanopore. The arrow indicates
the translocation direction.
(b) Cut-through view of the pore region for configuration in panel
(a). The chain backbone, with a tightened trefoil-knot  at the pore entrance, is highlighted in the inset.}
\end{figure}

We consider a model polyelectrolyte chain that is electrophoretically
driven through a pore embedded in the slab separating the {\it cis} and
{\it trans} semi-spaces \cite{zwolak_diventra}, see
Fig. \ref{fig:setup}. \textcolor{black}{For definiteness, the salient
  physical properties of the (otherwise general) model system are set to
  match those of 15$\mu$m-long ssDNA filaments in a solution with $0.1$M
  monovalent salt and translocated through an artificial nanopore. The
  nominal slab thickness and effective pore diameter were set respectively equal
  to 10nm and 2nm, consistent with the typical geometry of solid-state
  nanopores (which are about 3-4 times longer than biological
  nanopores) \cite{zwolak_diventra}.}

The polyelectrolyte chain consists of $N=15000$ beads, with diameter
equal to the nominal ssDNA thickness, $\sigma=1$nm.
\textcolor{black}{Because the monovalent counterions reduce the phosphates
  electrostatic charge by $\sim 50\%$ \cite{PhysRevLett.105.158101} the
    effective charge of the beads, each spanning three nucleotides, is
    set equal to $q=-1.5 e$}. The chain potential energy is accordingly
  given by:

\begin{equation}
{\cal H}=  \sum_{i=1}^N\{U_{\rm \small FENE}(d_{i,i+1}) + {1 \over 2} \sum_{j=1 \; j\not=i}^N \left[ U_{\rm LJ}(d_{i, j}) + U_{\rm DH}(d_{i, j}) \right ] \} \nonumber
\end{equation}
\noindent where $d_{i,j}$ is the distance of monomers $i$ and $j$, and the three terms,
which enforce respectively the chain connectivity constraint ($U_{\rm FENE}$),
the pairwise Lennard-Jones interaction ($U_{\rm LJ}$) and Debye-Hueckel ($U_{\rm DH}$) repulsion have the form:
\begin{eqnarray}
&&U_{\rm FENE}(d)=\left\{
\begin{array}{l}
- 150 \kappa_B\, T\, ({R_0 \over \sigma})^2  \, \ln \left[ 1 - \left( {d \over R_0} \right)^2 \right], \, d \leq R_0\\
0, \, d > R_0 
\end{array}
\right. \nonumber \\
&&U_{\rm LJ}(d)=\left\{
\begin{array}{l}
4 \kappa_B\, T \epsilon_{i,j} [({\sigma \over d})^{12} - ({\sigma \over d})^6 + 1/4], d \leq \sigma 2^{1/6}\\
0, d > \sigma\, 2^{1/6} \nonumber
\end{array}
\right. \label{eqn:LJ}\\
&&U_{\rm DH}(d)=\frac{q^2 \, e^{-d / \lambda_{\rm DH}}}{\varepsilon \, d}, \nonumber
\end{eqnarray}
\noindent where $R_0=1.5\sigma$, $\kappa_B$ is the Boltzmann constant,
$T=300$K, $\varepsilon=80$ is the water dielectric constant,
$\lambda_{\rm DH}=0.9$nm is the Debye screening length, and
$\epsilon_{i,j}$ is equal to 1 if $|i-j|=1$, and 10 otherwise.

A Monte Carlo scheme, employing unrestricted, non topology-preserving
local and global moves \cite{Madras_Sokal:1988:JSP,micheletti_physrep},
was first used to generate an equilibrated set of conformations for
flexible self-avoiding chains of $N=15000$ beads in bulk.  The degree and type of
entanglement of the filaments was established using the
minimally-interfering closure scheme
\cite{tubiana:2011:PTP,Tubiana:2011:Phys-Rev-Lett:22107680}.  It was
found that $\sim 2$\% of the configurations were knotted and, in more
than 90\% of the cases, they consisted of the simplest knot type: the
trefoil or $3_1$ knot (see inset in Fig. \ref{fig:setup}b for a trefoil
knot representation). The knots spanned, on average, about 10\% of the
chain; the sizable knot length is consistent with the expected
delocalisation of knots in unconstrained chains
\cite{marcone2007,virnau_kantor,mansfdoug2010}.

\textcolor{black}{Since the main focus of the Letter concerns the impact of chain topology on the translocation dynamics, we neglect here the otherwise important issue of how a chain in the bulk approaches the pore and enters it \cite{Muthukumar:2010:J-Chem-Phys:20499989}. Accordingly,} we extracted from the equilibrated ensemble several uncorrelated knotted and unknotted chains,
for which a rigid global translation or rotation
could bring one of the ends at the pore entrance while the chain remainder
stays in the {\it cis} semispace.

The beads inside the pore are driven
through it by a constant force whose magnitude, $f$, typically chosen in
the $4-40$ pN range. \textcolor{black}{Assuming that the electro-osmotic
  screening inside the
  channel \cite{PhysRevE.78.021912,Electrophoretic_DNA_Translocation}
  reduces by $\sim 50\%$ the charge density of ssDNA,
  these forces correspond to a uniform electric field of $0.15-1.5$ V
  per 10nm acting on each bead.}  The translocation dynamics \textcolor{black}{of the polyelectrolyte chain} is
integrated numerically using the fixed-volume and constant-temperature
molecular dynamics simulation scheme implemented in the LAMMPS package
\cite{lammps}. As in other coarse-grained approaches, no explicit hydrodynamic treatment is introduced \cite{ali_yeomans}.

\begin{figure}
\includegraphics[width=3.0in]{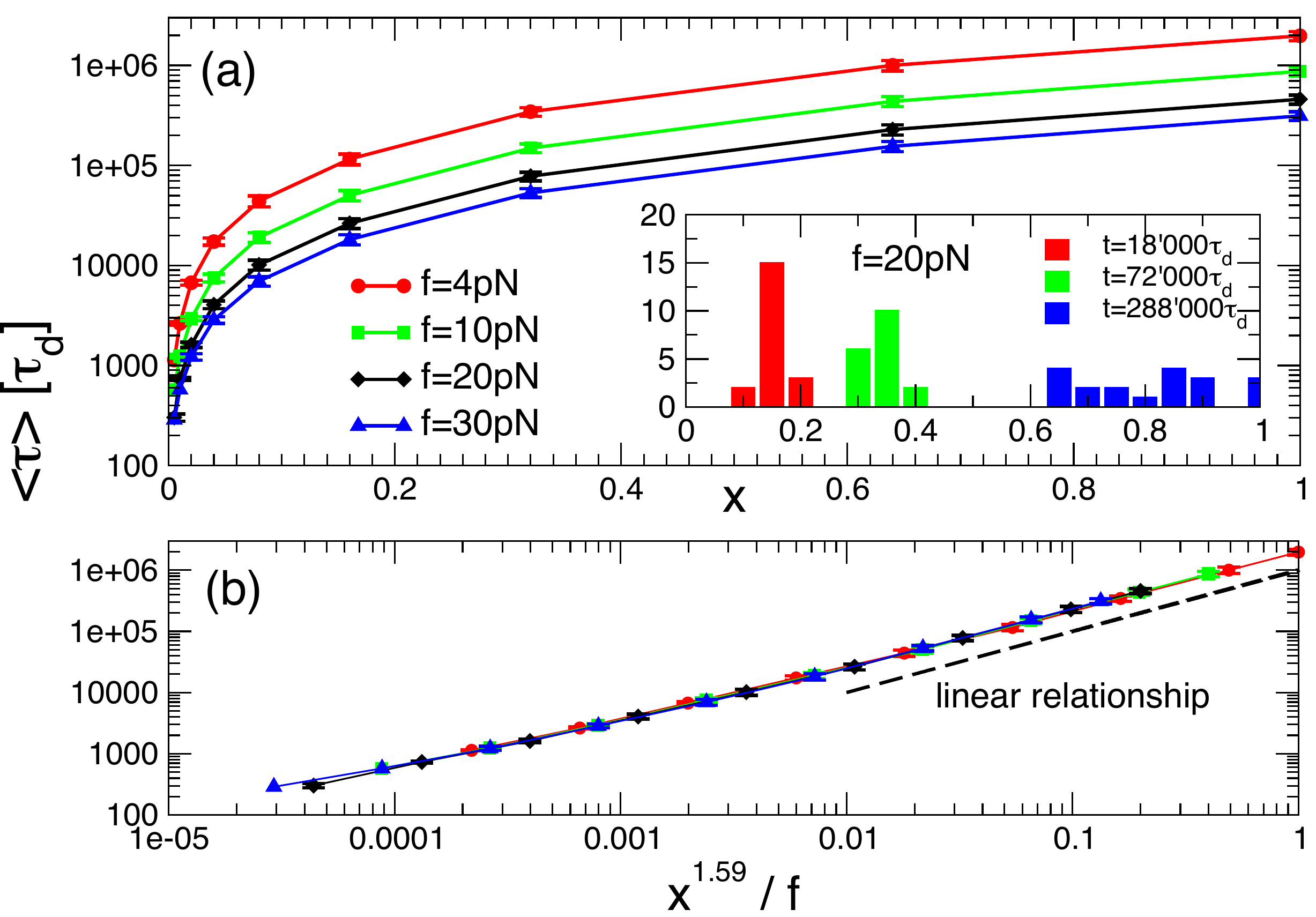}
\caption{
[Color online]
(a) Average time, $\langle \tau \rangle $, required to
  translocate a fraction $x$ of an unknotted chain at various driving
  forces, $f$. The average is taken over 20 uncorrelated unknotted
  conformations. Inset: statistical distribution of $x$
  at three different times and $f=20$pN.  Because
  the process timescale falls in the overdamped regime, time is
  expressed in units of the nominal monomer self-diffusion time: $\tau_d
  = \pi \eta \sigma^3 / 2 \kappa_B T$. For water viscosity, $\eta=$ 1cP,
  $\tau_d\sim 0.5$ns \cite{kremer_jcp}.  (b) Collapse of the data-points
  in panel (a) using the theoretical dynamical scaling, see text.
\label{fig:unknots}}
\end{figure}

{\it Unknotted chain dynamics -} The translocation dynamics for various
pulling forces and averaged over an ensemble of 20 {\it unknotted}
chains is illustrated in Fig. \ref{fig:unknots}a.  The dynamical process
is sensitive to the specific geometrical arrangements of the pulled
chains: in fact, the heterogeneity of the process increases with time.
As shown in the inset at fixed force and at a given simulation time, the
translocated fraction of the chain can range from 50\% to 100\%
depending on the initial configuration.  Notwithstanding these
differences, the average asymptotic translocation times appear to follow
closely the dynamical scaling relationship predicted \textcolor{black}{theoretically,
$\langle \tau \rangle \propto x^{1+ \nu}/f$}
where $x$ is the translocated fraction of the chain
\cite{PhysRevE.69.021806,PhysRevE.78.021129,grosberg_transloc}.
In fact, by using the theoretical self-avoiding matrix exponent $\nu=0.59$, we find the above expression
well satisfied, see Fig.~\ref{fig:unknots}b.

It should also be noted that this scaling relationship, which was first
formulated using dimensional and heuristic arguments
\cite{PhysRevE.69.021806,PhysRevE.78.021129,grosberg_transloc}, \textcolor{black}{was more recently shown to hold only for asymptotically long chains \cite{PhysRevE.85.051803}. In fact the non-equilibrium process that governs the  propagation of the tensile disturbance along the chain is sensitive to finite chain effects} \cite{PhysRevE.85.051803}.  The
collapse of the data points in Fig.~\ref{fig:unknots}b clarify that the
theoretically-predicted \textcolor{black}{asymptotic scaling relationship holds satisfactorily for chains 
of $N=15000$ beads with the considered  driving protocol}.

\begin{figure}
\includegraphics[width=2.8in]{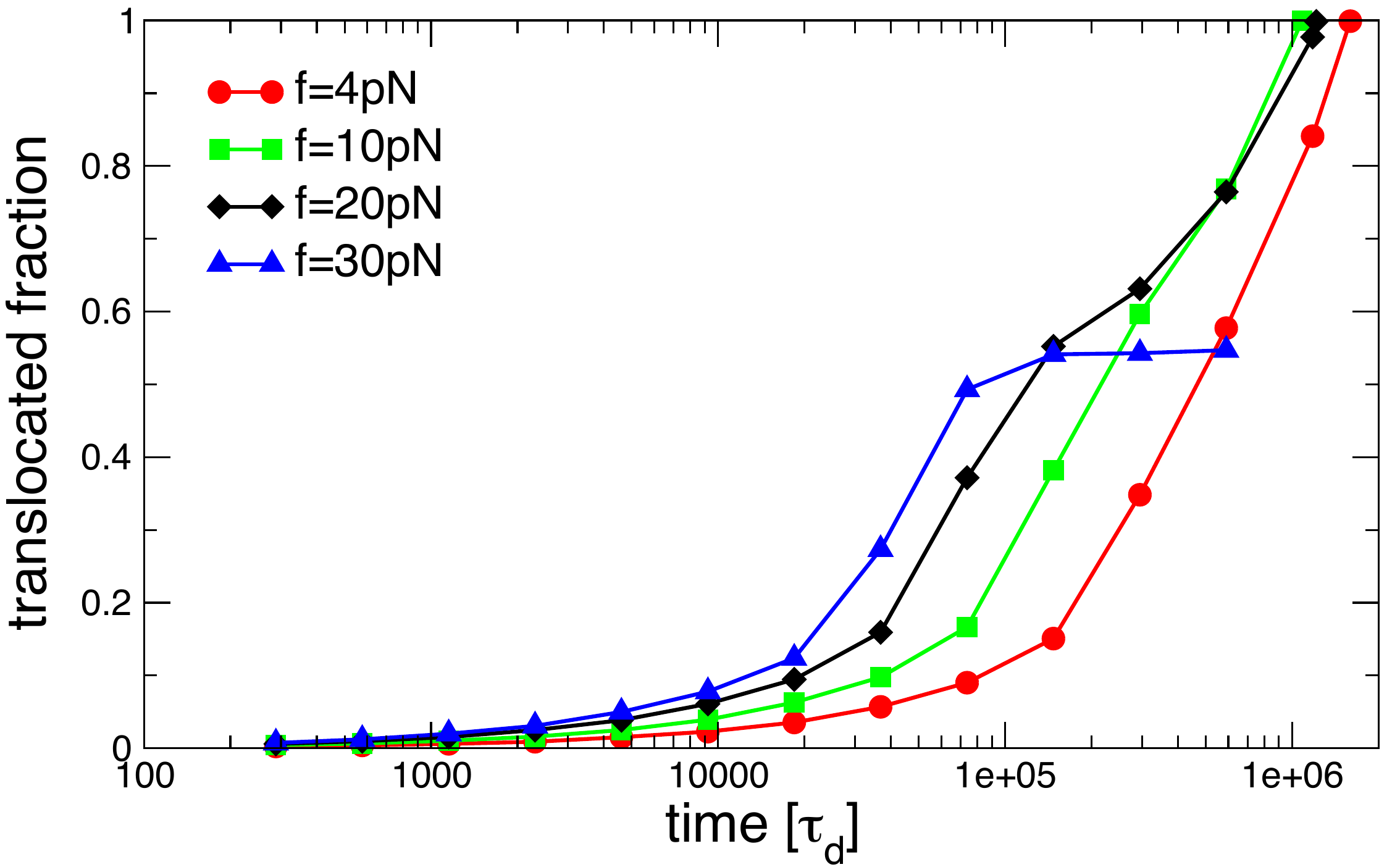}
\caption{
\label{fig:beadindex}
[Color online]
\textcolor{black}{Translocation kinetics of a specific trefoil-knotted chain at various driving forces.}}
\end{figure}

{\it Knotted chain dynamics -} Compared to the unknotted case, the
translocation dynamics of knotted chains has instead a dramatic,
non-monotonic dependence on the driving force.  This is illustrated in
Fig. \ref{fig:beadindex} \textcolor{black}{for one particular
  trefoil-knotted configuration} where the knot occupies the mid-portion
of the chain and spans about 10\% of the chain.  It is seen that, at the
smallest driving force the chain translocation is well-consistent with
the average translocation dynamics of unknotted chains.  At higher
forces, however, the standard dynamic scaling is satisfied only up to
when about $50$\% of the chain is translocated, after which a noticeable
slowing down of the process ensues.  Notice that at the highest force,
the translocation process appears to be practically halted.

\begin{figure}
\includegraphics[width=3.0in]{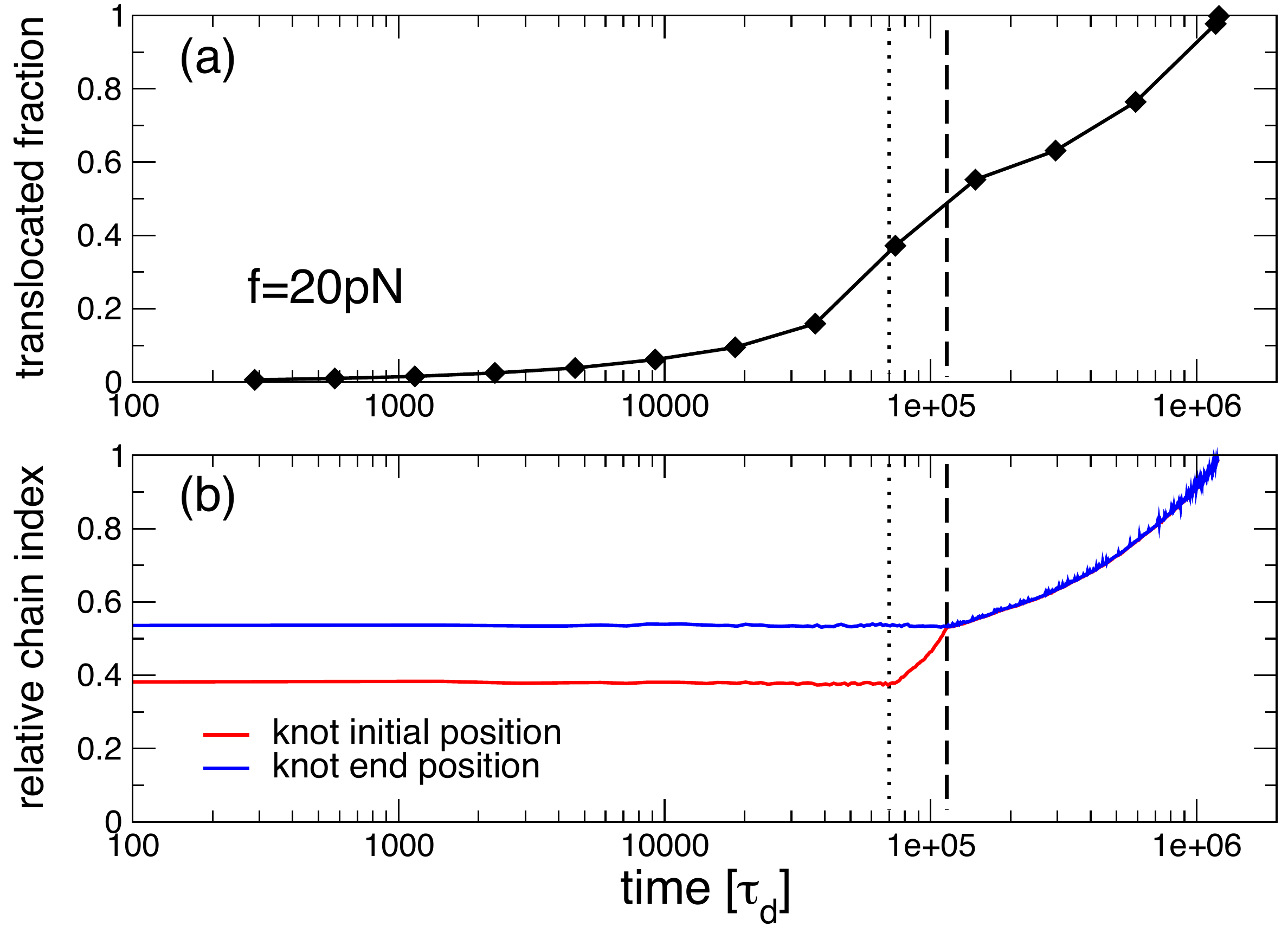}
\caption{
\label{fig:knotvelocity}
[Color online]
\textcolor{black}{Dynamical evolution (a) of the translocated chain
  fraction and (b) of the relative knot location along the chain for the
  same translocation run at $f=20pN$ shown in
  Fig.~\ref{fig:beadindex}. At time $t\sim 7\,10^4 \tau_{d}$ (dotted
  line) the propagating pulling front reaches the knot which
  progressively tightens until it reaches the pore at $t\approx 1.2\,
  10^5\tau_{d}$ (dashed line).}}
\end{figure}

This {\it dynamical crossover} is understood by monitoring the
instantaneous position and size of the knot along the chain.  The data
for $f=20$pN are shown in Fig.~\ref{fig:knotvelocity}.  It is seen that
the knot position along the chain is unperturbed until it is reached by
the propagating pulling front.  At this stage, the progressive dragging
of the chain through the pore causes the knot to tighten.  Notice that,
because the sequence index of the knot distal end remains about constant
in time, the knot tightening process consists of the removal of the
``slack'' from the end that is nearest to the pore.  After this stage,
the tightened knot is localized at the pore entrance and the
translocation proceeds by chain reptation through the knot
``defect''. These results illustrate the remarkable impact that
non-trivial {\it spontaneous entanglement} of long polymer chains has on
the translocation dynamics as a function of the pulling force.

To quantify in the most transparent way the topology-dependent aspect of
the effect, and separate it from the one associated to  the chain geometry
we have extended our analysis in two complementary directions. First, by
comparing the dynamics of chains with same geometry, but different
topology near the pore entrance. Secondly, by suitably averaging the
translocation dynamics over chains with different geometry but same knot
topology.

\begin{figure}
\includegraphics[width=3.0in]{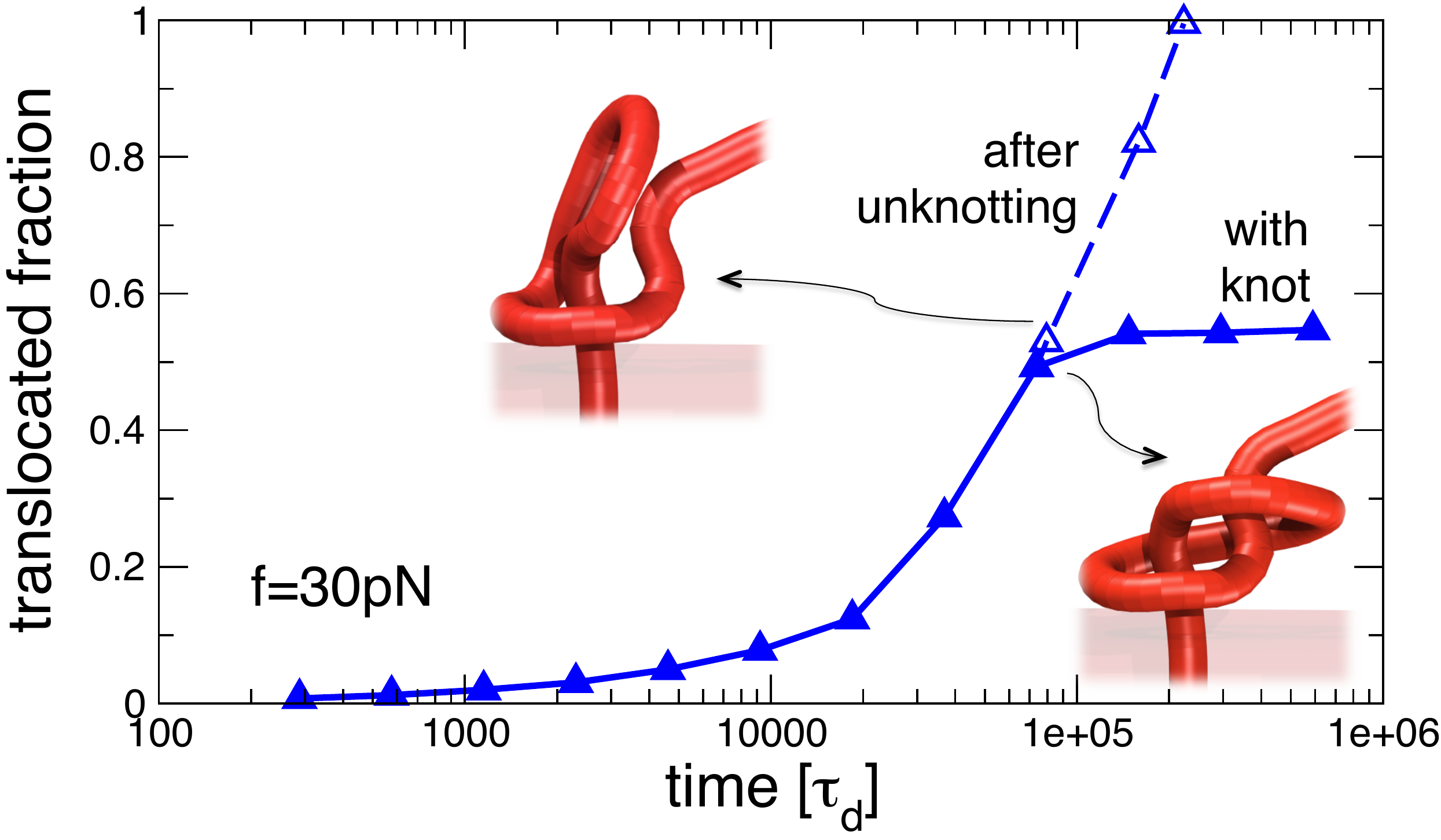}
\caption{
\label{fig:excision}
[Color online]
Dependence of the translocation dynamics on the chain topology. After
unknotting the chain by the shown local modification of the chain at the
pore entrance the translocation velocity is dramatically enhanced. For clarity, here and in Fig.~\ref{fig:friction} the chain centerline is rendered as a continuous tube with diameter smaller than $\sigma$.}
\end{figure}

For the first analysis, immediately after the knot is tightened at the
pore entrance, one can locally perturb the chain geometry at the pore so
as to untie the knot, while leaving unaltered the coordinates of all
other parts of the chains, see Fig.~\ref{fig:excision}a.  Next, by
following both the dynamical evolution of the knotted and unknotted
versions of the chain (with the same initial velocities for both
simulations) it is possible to compare the net effect of the localized
knot defect on the chain translocation dynamics.  The results are shown
in Fig.~\ref{fig:excision} and aptly illustrate that the slowing down of
the translocation process is ascribable to the presence of the localized
knot at the pore entrance.

\begin{figure}
\includegraphics[width=3.0in]{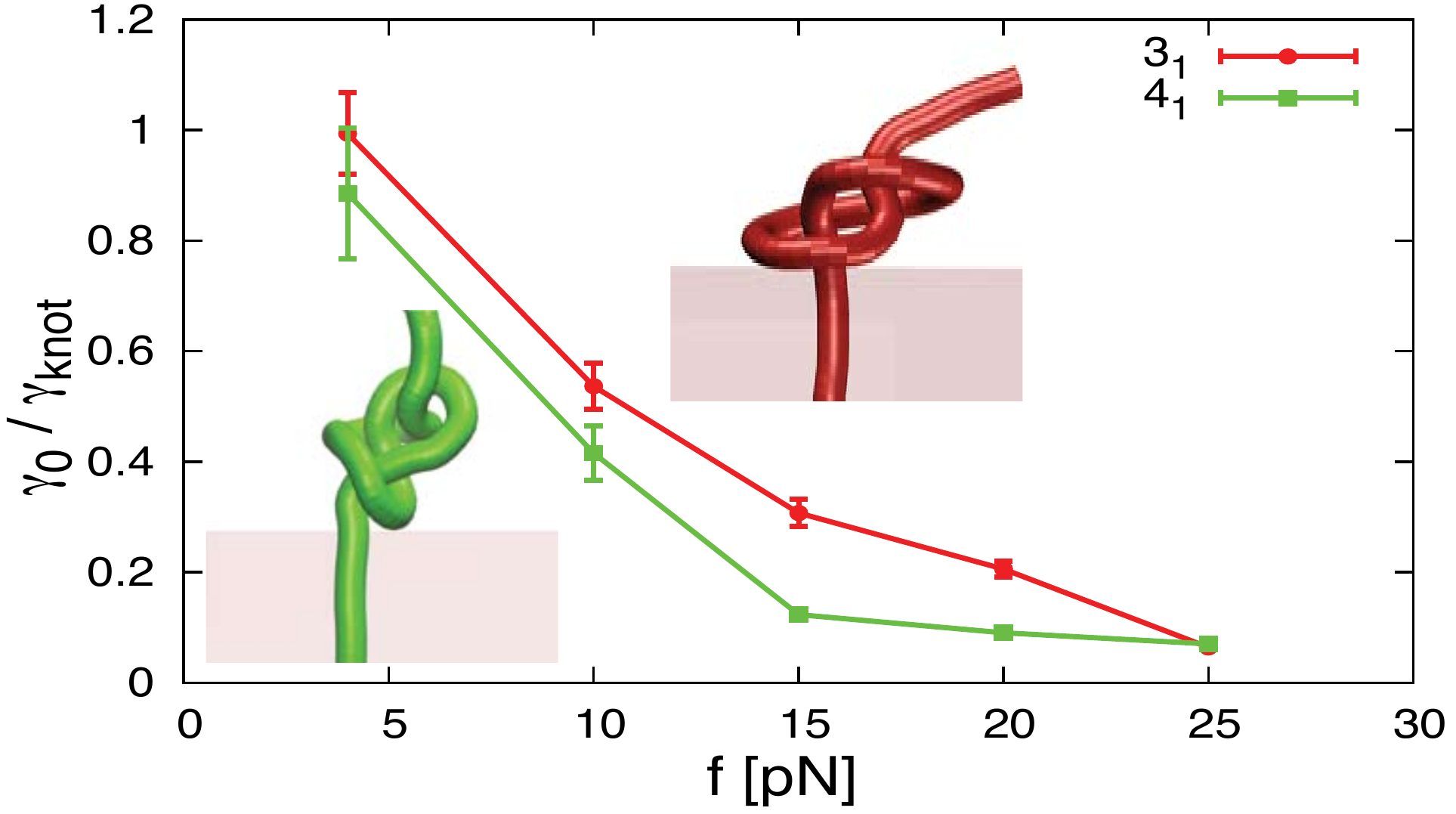}
\caption{
[Color online]
Force dependence of the normalised inverse friction coefficient,
$\gamma_0/\gamma_{knot}$ for chains hosting $3_1$ knots and $4_1$ knots. \textcolor{black}{At each force, $\gamma_0$ is defined as $\gamma_0 \equiv \langle \tau_t \rangle f /\sigma$, where $\langle \tau_t \rangle $ is the average time required by unknotted chains to fully translocate.} An analogous definition is used to compute $\gamma_{knot}$ for $3_1$- and $4_1$-knotted chains. Averages are taken over 20 configurations for the unknotted case and five configurations for each non-trivial topology.}
\label{fig:friction}
\end{figure}

A quantitative assessment of the topology-dependent hindrance can be
made by comparing the average translocation times of chains that are
unknotted (see Fig. 2a) and of chains whose knot is initially close to
the chain end at the pore entrance. Specifically, we considered chains
where the knot was located within the first 10-20\% of the chain, and for which
the knot tightening and trapping occurs within a time span that is typically less 
than 5\% of the full translocation process.  We computed the
effective friction coefficient, $\gamma_{knot}$, of these knot-dominated
processes and found that it has a dramatic dependence on $f$.  This
contrasts with the unentangled case where $\gamma_0$ is practically
force-independent, as testified by the collapse of the data in
Fig.~\ref{fig:unknots}.  The effect is illustrated in
Fig.~\ref{fig:friction}.

Notice that at the lowest pulling force, $f=4$pN, where a fairly tight
(spanning 27 monomers on average) knot is trapped at the pore entrance,
$\gamma_{knot}$ is statistically compatible with $\gamma_0$. We
therefore conclude that knot localization at the pore entrance is not
{\it per se} an impediment for translocation.  It becomes so only when
higher pulling forces cause the chain beads to interact very tightly,
resulting in a rapidly increasing effective friction. Consistent with
the dynamics in Figs. \ref{fig:knotvelocity} and \ref{fig:excision}, the
trend in Fig.~\ref{fig:friction} indicates that the translocation
process is practically halted at $f\sim 30$pN.

\textcolor{black}{It is interesting to notice that the observed impact of
  topology on dynamics differs from the case of knotted chains that are passively ejected  out of a small spherical cavity \cite{yeomansknots,micheletti_pnas}. In such systems, knots - even when tight - reduce the ejection speed by only a factor of $2-3$  \cite{yeomansknots}, possibly due to the relatively small
  magnitude of the force driving the spontaneous ejection.}

\textcolor{black}{Yet, consistent with available numerical results for knot-controlled DNA ejection~\cite{yeomansknots} and protein translocation dynamics~\cite{Huang:2008:J-Chem-Phys:19044999}  we do observe that at moderate driving forces, the
  dynamical hindrance is appreciably higher for knots that are more complex (and rarer) than trefoils.  In fact, as
  it is shown in Fig. ~\ref{fig:friction} the average inverse friction
  coefficient of five $4_1$-knotted chains is noticeably smaller than
  of $3_1$-knotted ones. Forces in the $30-40$pN range are nevertheless sufficient to
  halt the translocation process of all such $4_1$-knotted
  configurations as well as of five instances of $5_1$- and
  $5_2$-knotted ones (see supporting material).}

Finally, we stress that the results presented here ought to be relevant
in applicative contexts, such as genomic nanopore sequencing where the
high-throughput demand pushes towards interrogating longer and longer
uninterrupted DNA filaments at pulling forces comparable to those
considered in this work \cite{zwolak_diventra,branton08}.  The
high-throughput condition inevitably leads to significant chain
self-entanglement, while high pulling forces may cause tight knots to
halt the translocation process.  It should also be pointed out that in
actual ssDNA chains the latter effect will expectedly be more severe
than in our model system because of both the ramified character of the
molecule, and base-pairing effects. Since, the predicted force-dependent
topological jamming is experimentally verifiable, we hope that the
present investigation will motivate further studies aimed at quantifying
this effect.

We acknowledge support from the Italian Ministry of Education and the National Human Genome Research
Institute of NIH.


\begin{thebibliography}{41}
\expandafter\ifx\csname natexlab\endcsname\relax\def\natexlab#1{#1}\fi
\expandafter\ifx\csname bibnamefont\endcsname\relax
  \def\bibnamefont#1{#1}\fi
\expandafter\ifx\csname bibfnamefont\endcsname\relax
  \def\bibfnamefont#1{#1}\fi
\expandafter\ifx\csname citenamefont\endcsname\relax
  \def\citenamefont#1{#1}\fi
\expandafter\ifx\csname url\endcsname\relax
  \def\url#1{\texttt{#1}}\fi
\expandafter\ifx\csname urlprefix\endcsname\relax\def\urlprefix{URL }\fi
\providecommand{\bibinfo}[2]{#2}
\providecommand{\eprint}[2][]{\url{#2}}

\bibitem[{\citenamefont{Zwolak and {Di Ventra}}(2008)}]{zwolak_diventra}
\bibinfo{author}{\bibfnamefont{M.}~\bibnamefont{Zwolak}} \bibnamefont{and}
  \bibinfo{author}{\bibfnamefont{M.}~\bibnamefont{{Di Ventra}}},
  \bibinfo{journal}{Rev. Mod. Phys.} \textbf{\bibinfo{volume}{80}},
  \bibinfo{pages}{141} (\bibinfo{year}{2008}).

\bibitem[{\citenamefont{Branton et~al.}(2008)}]{branton08}
\bibinfo{author}{\bibfnamefont{D.}~\bibnamefont{Branton}} \bibnamefont{et~al.},
  \bibinfo{journal}{Nat. Biotechnol.} \textbf{\bibinfo{volume}{26}},
  \bibinfo{pages}{1146} (\bibinfo{year}{2008}).

\bibitem[{\citenamefont{Reisner et~al.}(2005)}]{austin_frey}
\bibinfo{author}{\bibfnamefont{W.}~\bibnamefont{Reisner}} \bibnamefont{et~al.},
  \bibinfo{journal}{Phys. Rev. Lett.} \textbf{\bibinfo{volume}{94}},
  \bibinfo{pages}{196101} (\bibinfo{year}{2005}).

\bibitem[{\citenamefont{Storm et~al.}(2005)}]{dekker_joanny}
\bibinfo{author}{\bibfnamefont{A.~J.} \bibnamefont{Storm}}
  \bibnamefont{et~al.}, \bibinfo{journal}{Nano Letters}
  \textbf{\bibinfo{volume}{5}}, \bibinfo{pages}{1193} (\bibinfo{year}{2005}).

\bibitem[{\citenamefont{Luan and
  Aksimentiev}(2010)}]{Luan:2010:Soft-Matter:20563230}
\bibinfo{author}{\bibfnamefont{B.}~\bibnamefont{Luan}} \bibnamefont{and}
  \bibinfo{author}{\bibfnamefont{A.}~\bibnamefont{Aksimentiev}},
  \bibinfo{journal}{Soft Matter} \textbf{\bibinfo{volume}{6}},
  \bibinfo{pages}{243} (\bibinfo{year}{2010}).

\bibitem[{\citenamefont{van Dorp
  et~al.}(2009)}]{Electrophoretic_DNA_Translocation}
\bibinfo{author}{\bibfnamefont{S.}~\bibnamefont{van Dorp}}
  \bibnamefont{et~al.}, \bibinfo{journal}{Nat. Phys.}
  \textbf{\bibinfo{volume}{5}}, \bibinfo{pages}{347} (\bibinfo{year}{2009}).

\bibitem[{\citenamefont{Kasianowicz et~al.}(1996)}]{Kasianowicz1996-1}
\bibinfo{author}{\bibfnamefont{J.~J.} \bibnamefont{Kasianowicz}}
  \bibnamefont{et~al.}, \bibinfo{journal}{Proc. Natl. Acad. Sci. USA}
  \textbf{\bibinfo{volume}{93}}, \bibinfo{pages}{13770} (\bibinfo{year}{1996}).

\bibitem[{\citenamefont{Lagerqvist et~al.}(2006)\citenamefont{Lagerqvist,
  Zwolak, and Di~Ventra}}]{Lagerqvist2006-1}
\bibinfo{author}{\bibfnamefont{J.}~\bibnamefont{Lagerqvist}},
  \bibinfo{author}{\bibfnamefont{M.}~\bibnamefont{Zwolak}}, \bibnamefont{and}
  \bibinfo{author}{\bibfnamefont{M.}~\bibnamefont{Di~Ventra}},
  \bibinfo{journal}{Nano Lett.} \textbf{\bibinfo{volume}{6}},
  \bibinfo{pages}{779} (\bibinfo{year}{2006}).

\bibitem[{\citenamefont{{D. W. Sumners, S. G.
  Whittington}}({1988})}]{Sumners&Whittington:1988:J-Phys-A}
\bibinfo{author}{\bibnamefont{{D. W. Sumners, S. G. Whittington}}},
  \bibinfo{journal}{{J. Phys. A: Math. Gen.}} \textbf{\bibinfo{volume}{{21}}},
  \bibinfo{pages}{1689} (\bibinfo{year}{{1988}}).

\bibitem[{\citenamefont{Ercolini et~al.}(2007)}]{Ercolini:2007:PRL}
\bibinfo{author}{\bibfnamefont{E.}~\bibnamefont{Ercolini}}
  \bibnamefont{et~al.}, \bibinfo{journal}{Phys. Rev. Lett.}
  \textbf{\bibinfo{volume}{98}}, \bibinfo{pages}{058102}
  (\bibinfo{year}{2007}).

\bibitem[{\citenamefont{Marenduzzo et~al.}(2010)\citenamefont{Marenduzzo,
  Micheletti, and Orlandini}}]{Marenduzzo_et_al_2010_JoP_Cond_Matt}
\bibinfo{author}{\bibfnamefont{D.}~\bibnamefont{Marenduzzo}},
  \bibinfo{author}{\bibfnamefont{C.}~\bibnamefont{Micheletti}},
  \bibnamefont{and}
  \bibinfo{author}{\bibfnamefont{E.}~\bibnamefont{Orlandini}},
  \bibinfo{journal}{J. Phys.: Condens. Matter} \textbf{\bibinfo{volume}{22}},
  \bibinfo{pages}{283102} (\bibinfo{year}{2010}).

\bibitem[{\citenamefont{Rybenkov et~al.}(1993)\citenamefont{Rybenkov,
  Cozzarelli, and Vologodskii}}]{rybenkov_cozzarelli}
\bibinfo{author}{\bibfnamefont{V.~V.} \bibnamefont{Rybenkov}},
  \bibinfo{author}{\bibfnamefont{N.~R.} \bibnamefont{Cozzarelli}},
  \bibnamefont{and} \bibinfo{author}{\bibfnamefont{A.~V.}
  \bibnamefont{Vologodskii}}, \bibinfo{journal}{Proc. Natl. Acad. Sci. USA}
  \textbf{\bibinfo{volume}{90}}, \bibinfo{pages}{5307} (\bibinfo{year}{1993}).

\bibitem[{\citenamefont{Saitta et~al.}(1999)}]{saitta_klein}
\bibinfo{author}{\bibfnamefont{A.~M.} \bibnamefont{Saitta}}
  \bibnamefont{et~al.}, \bibinfo{journal}{Nature}
  \textbf{\bibinfo{volume}{399}}, \bibinfo{pages}{46} (\bibinfo{year}{1999}).

\bibitem[{\citenamefont{Liu et~al.}(2009)}]{chan_zechiedrich}
\bibinfo{author}{\bibfnamefont{Z.~R.} \bibnamefont{Liu}} \bibnamefont{et~al.},
  \bibinfo{journal}{Nucleic Acids Res.} \textbf{\bibinfo{volume}{37}},
  \bibinfo{pages}{661} (\bibinfo{year}{2009}).

\bibitem[{\citenamefont{Deguchi and
  Tsurusaki}(1993)}]{Deguchi&Tsurusaki:1993:J-Phys-Soc-Jap}
\bibinfo{author}{\bibfnamefont{T.}~\bibnamefont{Deguchi}} \bibnamefont{and}
  \bibinfo{author}{\bibfnamefont{K.}~\bibnamefont{Tsurusaki}},
  \bibinfo{journal}{J. Phys. Soc. Japan} \textbf{\bibinfo{volume}{62}},
  \bibinfo{pages}{1411} (\bibinfo{year}{1993}).

\bibitem[{\citenamefont{Bao et~al.}(2003)\citenamefont{Bao, Lee, and
  Quake}}]{Bao:2003:Phys-Rev-Lett:14754067}
\bibinfo{author}{\bibfnamefont{X.~R.} \bibnamefont{Bao}},
  \bibinfo{author}{\bibfnamefont{H.~J.} \bibnamefont{Lee}}, \bibnamefont{and}
  \bibinfo{author}{\bibfnamefont{S.~R.} \bibnamefont{Quake}},
  \bibinfo{journal}{Phys. Rev. Lett.} \textbf{\bibinfo{volume}{91}},
  \bibinfo{pages}{265506} (\bibinfo{year}{2003}).

\bibitem[{\citenamefont{Sulkowska
  et~al.}(2008)}]{Sulkowska:2008:Phys-Rev-Lett:18352439}
\bibinfo{author}{\bibfnamefont{J.~I.} \bibnamefont{Sulkowska}}
  \bibnamefont{et~al.}, \bibinfo{journal}{Phys. Rev. Lett.}
  \textbf{\bibinfo{volume}{100}}, \bibinfo{pages}{058106}
  (\bibinfo{year}{2008}).

\bibitem[{\citenamefont{Huang and
  Makarov}(2008)}]{Huang:2008:J-Chem-Phys:19044999}
\bibinfo{author}{\bibfnamefont{L.}~\bibnamefont{Huang}} \bibnamefont{and}
  \bibinfo{author}{\bibfnamefont{D.~E.} \bibnamefont{Makarov}},
  \bibinfo{journal}{J. Chem. Phys.} \textbf{\bibinfo{volume}{129}},
  \bibinfo{pages}{121107} (\bibinfo{year}{2008}).

\bibitem[{\citenamefont{Meluzzi et~al.}(2010)\citenamefont{Meluzzi, Smith, and
  Arya}}]{Meluzzi:2010:Annu-Rev-Biophys:20192771}
\bibinfo{author}{\bibfnamefont{D.}~\bibnamefont{Meluzzi}},
  \bibinfo{author}{\bibfnamefont{D.~E.} \bibnamefont{Smith}}, \bibnamefont{and}
  \bibinfo{author}{\bibfnamefont{G.}~\bibnamefont{Arya}},
  \bibinfo{journal}{Annu. Rev. Biophys.} \textbf{\bibinfo{volume}{39}},
  \bibinfo{pages}{349} (\bibinfo{year}{2010}).

\bibitem[{\citenamefont{Kantor and Kardar}(2004)}]{PhysRevE.69.021806}
\bibinfo{author}{\bibfnamefont{Y.}~\bibnamefont{Kantor}} \bibnamefont{and}
  \bibinfo{author}{\bibfnamefont{M.}~\bibnamefont{Kardar}},
  \bibinfo{journal}{Phys. Rev. E} \textbf{\bibinfo{volume}{69}},
  \bibinfo{pages}{021806} (\bibinfo{year}{2004}).

\bibitem[{\citenamefont{Chatelain et~al.}(2008)\citenamefont{Chatelain, Kantor,
  and Kardar}}]{PhysRevE.78.021129}
\bibinfo{author}{\bibfnamefont{C.}~\bibnamefont{Chatelain}},
  \bibinfo{author}{\bibfnamefont{Y.}~\bibnamefont{Kantor}}, \bibnamefont{and}
  \bibinfo{author}{\bibfnamefont{M.}~\bibnamefont{Kardar}},
  \bibinfo{journal}{Phys. Rev. E} \textbf{\bibinfo{volume}{78}},
  \bibinfo{pages}{021129} (\bibinfo{year}{2008}).

\bibitem[{\citenamefont{Grosberg et~al.}(2006)}]{grosberg_transloc}
\bibinfo{author}{\bibfnamefont{A.~Y.} \bibnamefont{Grosberg}}
  \bibnamefont{et~al.}, \bibinfo{journal}{Phys. Rev. Lett.}
  \textbf{\bibinfo{volume}{96}}, \bibinfo{pages}{228105}
  (\bibinfo{year}{2006}).

\bibitem[{\citenamefont{Sakaue}(2007)}]{sakauePRE}
\bibinfo{author}{\bibfnamefont{T.}~\bibnamefont{Sakaue}},
  \bibinfo{journal}{Phys. Rev. E} \textbf{\bibinfo{volume}{76}},
  \bibinfo{pages}{021803} (\bibinfo{year}{2007}).

\bibitem[{\citenamefont{Sakaue and Rapha{\"e}l}(2006)}]{sakaue_raphael}
\bibinfo{author}{\bibfnamefont{T.}~\bibnamefont{Sakaue}} \bibnamefont{and}
  \bibinfo{author}{\bibfnamefont{E.}~\bibnamefont{Rapha{\"e}l}},
  \bibinfo{journal}{Macromolecules} \textbf{\bibinfo{volume}{39}},
  \bibinfo{pages}{2621} (\bibinfo{year}{2006}).

\bibitem[{\citenamefont{Forrey and Muthukumar}(2007)}]{forrey_muthukumar}
\bibinfo{author}{\bibfnamefont{C.}~\bibnamefont{Forrey}} \bibnamefont{and}
  \bibinfo{author}{\bibfnamefont{M.}~\bibnamefont{Muthukumar}},
  \bibinfo{journal}{J. Chem. Phys.} \textbf{\bibinfo{volume}{127}},
  \bibinfo{pages}{015102} (\bibinfo{year}{2007}).

\bibitem[{\citenamefont{Ikonen et~al.}(2012)\citenamefont{Ikonen, Bhattacharya,
  Ala-Nissila, and Sung}}]{PhysRevE.85.051803}
\bibinfo{author}{\bibfnamefont{T.}~\bibnamefont{Ikonen}},
  \bibinfo{author}{\bibfnamefont{A.}~\bibnamefont{Bhattacharya}},
  \bibinfo{author}{\bibfnamefont{T.}~\bibnamefont{Ala-Nissila}},
  \bibnamefont{and} \bibinfo{author}{\bibfnamefont{W.}~\bibnamefont{Sung}},
  \bibinfo{journal}{Phys. Rev. E} \textbf{\bibinfo{volume}{85}},
  \bibinfo{pages}{051803} (\bibinfo{year}{2012}).

\bibitem[{\citenamefont{Maffeo et~al.}(2010)}]{PhysRevLett.105.158101}
\bibinfo{author}{\bibfnamefont{C.}~\bibnamefont{Maffeo}} \bibnamefont{et~al.},
  \bibinfo{journal}{Phys. Rev. Lett.} \textbf{\bibinfo{volume}{105}},
  \bibinfo{pages}{158101} (\bibinfo{year}{2010}).

\bibitem[{\citenamefont{{N. Madras, A. D.
  Sokal}}(1988)}]{Madras_Sokal:1988:JSP}
\bibinfo{author}{\bibnamefont{{N. Madras, A. D. Sokal}}}, \bibinfo{journal}{J.
  Stat. Phys.} \textbf{\bibinfo{volume}{50}}, \bibinfo{pages}{109}
  (\bibinfo{year}{1988}).

\bibitem[{\citenamefont{Micheletti et~al.}(2011)\citenamefont{Micheletti,
  Marenduzzo, and Orlandini}}]{micheletti_physrep}
\bibinfo{author}{\bibfnamefont{C.}~\bibnamefont{Micheletti}},
  \bibinfo{author}{\bibfnamefont{D.}~\bibnamefont{Marenduzzo}},
  \bibnamefont{and}
  \bibinfo{author}{\bibfnamefont{E.}~\bibnamefont{Orlandini}},
  \bibinfo{journal}{Phys. Reports} \textbf{\bibinfo{volume}{504}},
  \bibinfo{pages}{1} (\bibinfo{year}{2011}).

\bibitem[{\citenamefont{Tubiana
  et~al.}(2011{\natexlab{a}})\citenamefont{Tubiana, Orlandini, and
  Micheletti}}]{tubiana:2011:PTP}
\bibinfo{author}{\bibfnamefont{L.}~\bibnamefont{Tubiana}},
  \bibinfo{author}{\bibfnamefont{E.}~\bibnamefont{Orlandini}},
  \bibnamefont{and}
  \bibinfo{author}{\bibfnamefont{C.}~\bibnamefont{Micheletti}},
  \bibinfo{journal}{Prog. Theor. Phys.} \textbf{\bibinfo{volume}{191}},
  \bibinfo{pages}{192} (\bibinfo{year}{2011}{\natexlab{a}}).

\bibitem[{\citenamefont{Tubiana
  et~al.}(2011{\natexlab{b}})\citenamefont{Tubiana, Orlandini, and
  Micheletti}}]{Tubiana:2011:Phys-Rev-Lett:22107680}
\bibinfo{author}{\bibfnamefont{L.}~\bibnamefont{Tubiana}},
  \bibinfo{author}{\bibfnamefont{E.}~\bibnamefont{Orlandini}},
  \bibnamefont{and}
  \bibinfo{author}{\bibfnamefont{C.}~\bibnamefont{Micheletti}},
  \bibinfo{journal}{Phys. Rev. Lett.} \textbf{\bibinfo{volume}{107}},
  \bibinfo{pages}{188302} (\bibinfo{year}{2011}{\natexlab{b}}).

\bibitem[{\citenamefont{Marcone et~al.}(2007)}]{marcone2007}
\bibinfo{author}{\bibfnamefont{B.}~\bibnamefont{Marcone}} \bibnamefont{et~al.},
  \bibinfo{journal}{Phys. Rev. E} \textbf{\bibinfo{volume}{75}},
  \bibinfo{pages}{041105} (\bibinfo{year}{2007}).

\bibitem[{\citenamefont{Virnau et~al.}(2005)\citenamefont{Virnau, Kantor, and
  Kardar}}]{virnau_kantor}
\bibinfo{author}{\bibfnamefont{P.}~\bibnamefont{Virnau}},
  \bibinfo{author}{\bibfnamefont{Y.}~\bibnamefont{Kantor}}, \bibnamefont{and}
  \bibinfo{author}{\bibfnamefont{M.}~\bibnamefont{Kardar}},
  \bibinfo{journal}{J. Am. Chem. Soc.} \textbf{\bibinfo{volume}{127}},
  \bibinfo{pages}{15102} (\bibinfo{year}{2005}).

\bibitem[{\citenamefont{Mansfield and Douglas}(2010)}]{mansfdoug2010}
\bibinfo{author}{\bibfnamefont{M.~L.} \bibnamefont{Mansfield}}
  \bibnamefont{and} \bibinfo{author}{\bibfnamefont{J.~F.}
  \bibnamefont{Douglas}}, \bibinfo{journal}{J. Chem. Phys.}
  \textbf{\bibinfo{volume}{133}}, \bibinfo{pages}{044903}
  (\bibinfo{year}{2010}).

\bibitem[{\citenamefont{Muthukumar}(2010)}]{Muthukumar:2010:J-Chem-Phys:204999%
89}
\bibinfo{author}{\bibfnamefont{M.}~\bibnamefont{Muthukumar}},
  \bibinfo{journal}{J Chem Phys} \textbf{\bibinfo{volume}{132}},
  \bibinfo{pages}{195101} (\bibinfo{year}{2010}).

\bibitem[{\citenamefont{Luan and Aksimentiev}(2008)}]{PhysRevE.78.021912}
\bibinfo{author}{\bibfnamefont{B.}~\bibnamefont{Luan}} \bibnamefont{and}
  \bibinfo{author}{\bibfnamefont{A.}~\bibnamefont{Aksimentiev}},
  \bibinfo{journal}{Phys. Rev. E} \textbf{\bibinfo{volume}{78}},
  \bibinfo{pages}{021912} (\bibinfo{year}{2008}).

\bibitem[{\citenamefont{Plimpton}(1995)}]{lammps}
\bibinfo{author}{\bibfnamefont{S.}~\bibnamefont{Plimpton}},
  \bibinfo{journal}{J. Comp. Phys.} \textbf{\bibinfo{volume}{117}},
  \bibinfo{pages}{119} (\bibinfo{year}{1995}).

\bibitem[{\citenamefont{Ali and Yeomans}(2005)}]{ali_yeomans}
\bibinfo{author}{\bibfnamefont{I.}~\bibnamefont{Ali}} \bibnamefont{and}
  \bibinfo{author}{\bibfnamefont{J.~M.} \bibnamefont{Yeomans}},
  \bibinfo{journal}{J. Chem. Phys.} \textbf{\bibinfo{volume}{123}},
  \bibinfo{pages}{234903} (\bibinfo{year}{2005}).

\bibitem[{\citenamefont{Kremer and Grest}(1990)}]{kremer_jcp}
\bibinfo{author}{\bibfnamefont{K.}~\bibnamefont{Kremer}} \bibnamefont{and}
  \bibinfo{author}{\bibfnamefont{G.~S.} \bibnamefont{Grest}},
  \bibinfo{journal}{J. Chem. Phys.} \textbf{\bibinfo{volume}{92}},
  \bibinfo{pages}{5057} (\bibinfo{year}{1990}).

\bibitem[{\citenamefont{Matthews et~al.}(2009)\citenamefont{Matthews, Louis,
  and Yeomans}}]{yeomansknots}
\bibinfo{author}{\bibfnamefont{R.}~\bibnamefont{Matthews}},
  \bibinfo{author}{\bibfnamefont{A.~A.} \bibnamefont{Louis}}, \bibnamefont{and}
  \bibinfo{author}{\bibfnamefont{J.~M.} \bibnamefont{Yeomans}},
  \bibinfo{journal}{Phys. Rev. Lett.} \textbf{\bibinfo{volume}{102}},
  \bibinfo{pages}{088101} (\bibinfo{year}{2009}).

\bibitem[{\citenamefont{Marenduzzo et~al.}(2009)}]{micheletti_pnas}
\bibinfo{author}{\bibfnamefont{D.}~\bibnamefont{Marenduzzo}}
  \bibnamefont{et~al.}, \bibinfo{journal}{Proc. Natl. Acad. Sci. USA}
  \textbf{\bibinfo{volume}{106}}, \bibinfo{pages}{22269}
  (\bibinfo{year}{2009}).

\end{thebibliography}

\setcounter{figure}{0}
\renewcommand{\figurename}{SUPPORTING FIG.}
\newpage

\begin{figure*}[h!]
\includegraphics[width=0.50\textwidth]{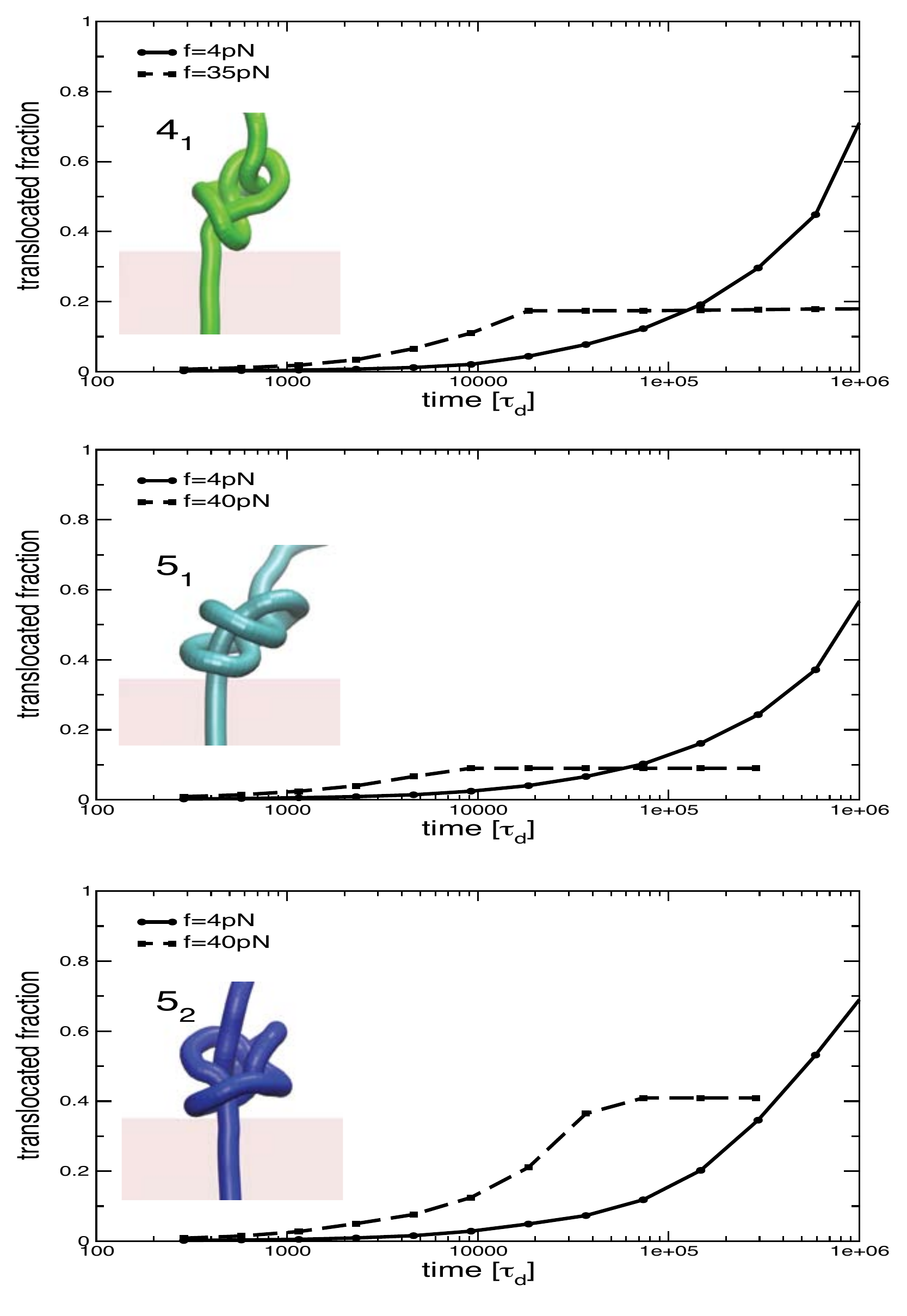}
\caption{
Translocation kinetics at low and high force for polymer chains with three different knot topologies $4_1$, $5_1$ and $5_2$ as shown in the insets.
At low force, $f=4$pN, the translocation proceeds steadily but it is stalled at $f\approx 35$pN for the $4_1$-knotted chain and
$f \approx 40$pN for $5_1$- and $5_2$-knotted chains.
The jamming force shown in the examples is typical: 5 instances of $4_1$ knots were all jammed at $f=35$pN and
5 instances of five-crossing knots were all jammed at $f=40$pN.
A time step as small as $0.003 \tau_{LJ}$,($\tau_{LJ} = \sigma \sqrt{m/\epsilon}$ is the Lennard-Jones time
associated to monomer dynamics \cite{kremer_jcp}) was used to avoid chain breaking  due to the build-up of tensile strain in jammed configurations.
The default LAMMPS parametrization \cite{kremer_jcp} $\eta = \frac{ \sqrt{m \epsilon} } {6 \pi \sigma^2}$ implies that
the effective bead mass, $m$ is $\approx 2.5 \cdot 10^{-19}$kg.
We recall in fact that
$\epsilon = \kappa_B T$ ($\kappa_B$ is the Boltzmann constant and $T=300$K),
$\sigma = 1$nm, and
$\eta = 1$cP.
With this effective bead mass, $\tau_{MD}$ corresponds to $\approx 6$ns.
By analysing the velocity-velocity autocorrelation function for internal chain beads,
it is established that the overdamped Langevin regime sets in at time-scales $\approx 0.5$ns,
i.e. comparable to $\tau_d$. In our analysis we accordingly  focused exclusively on the mass-independent MD evolution over time-spans longer than $\tau_d$ and have accordingly expressed time-scales in units of $\tau_d$.}
\end{figure*}

\end{document}